%% file: main.tex
\documentclass[sigconf]{acmart} 
\usepackage{subcaption}

\usepackage{algorithm}
\usepackage{multirow}
\usepackage{enumitem}
\usepackage{bm}
\usepackage{xspace}
\usepackage{marvosym}

\usepackage{bbding}
\AtBeginDocument{%
  }

\newcommand{\ie}{\emph{i.e.,}\xspace}
\newcommand{\eg}{\emph{e.g.,}\xspace}

\newcommand{\paratitle}[1]{\vspace{1.5ex}\noindent\textbf{#1}}
\newcommand{\wrt}{w.r.t.\xspace}

\newcommand{\ignore}[1]{}

\copyrightyear{2026}
\acmYear{2026}
\setcopyright{cc}
\setcctype{by}
\acmConference[WWW '26]{Proceedings of the ACM Web Conference 2026}{April 13--17, 2026}{Dubai, United Arab Emirates}
\acmBooktitle{Proceedings of the ACM Web Conference 2026 (WWW '26), April 13--17, 2026, Dubai, United Arab Emirates}
\acmPrice{}
\acmDOI{10.1145/3774904.3792612}
\acmISBN{979-8-4007-2307-0/2026/04}

\begin{document}

\title[ GenCI: Generative Modeling of User Interest Shift via Cohort-based Intent Learning for CTR Prediction]{\texorpdfstring{GenCI: Generative Modeling of User Interest Shift via  \\Cohort-based Intent Learning for CTR Prediction}{GenCI: Generative Modeling of User Interest Shift via Cohort-based Intent Learning for CTR Prediction}}

\settopmatter{authorsperrow=3}

\author{Kesha Ou}
\orcid{0009-0006-8557-5437}
\affiliation{
  \institution{Gaoling School of Artificial Intelligence, Renmin University of China}
  \city{Beijing}
  \country{China}
}
\authornote{Also with Beijing Key Laboratory of Research on Large Models and Intelligent Governance.}
\email{keishaou@gmail.com}

\author{Zhen Tian}
\orcid{0000-0001-5569-2591}
\affiliation{
    \institution{
    Gaoling School of Artificial Intelligence,  
    Renmin University of China}
    \city{Beijing}
    \country{China}
}
\email{chenyuwuxinn@gmail.com}

\author{Wayne Xin Zhao
%$\dagger$
\textsuperscript{\Letter}
}
\orcid{0000-0002-8333-6196}
\affiliation{
    \institution{
    Gaoling School of Artificial Intelligence,  
    Renmin University of China}
    \city{Beijing}
    \country{China}
}
\authornotemark[1]
\email{batmanfly@gmail.com}
\thanks{\Letter \ Corresponding author.}

\author{Hongyu Lu}
\orcid{0000-0002-0247-2496}
\affiliation{
 \institution{
 WeChat, Tencent
 }
     \city{Beijing}
 \country{China}
}
\email{luhy94@gmail.com}

\author{Ji-Rong Wen}
\orcid{0000-0002-9777-9676}
\affiliation{
    \institution{
    Gaoling School of Artificial Intelligence,  
    Renmin University of China}
    \city{Beijing}
    \country{China}
}
\authornotemark[1]
\email{jrwen@ruc.edu.cn}

\newcommand{\tba}{\textcolor{red}{xxx }}
\newcommand{\outd}{\textcolor{red}{[Outdated]}~}
\newcommand{\tabincell}[2]{\begin{tabular}{@{}#1@{}}#2\end{tabular}}

\definecolor{dark2green}{rgb}{0.1, 0.65, 0.3}
\definecolor{dark2orange}{rgb}{0.9, 0.4, 0.}
\definecolor{dark2purple}{rgb}{0.4, 0.4, 0.8}
\newcommand{\first}[1]{\textbf{#1}}
\newcommand{\second}[1]{\underline{#1}}
\newcommand{\third}[1]{\textbf{\textcolor{dark2purple}{#1}}}

\begin{abstract}

Click-through rate (CTR) prediction plays a pivotal role in online advertising and recommender systems. Despite notable progress in modeling user preferences from historical behaviors, two key challenges persist.  First, exsiting discriminative paradigms focus on matching candidates to user history, often overfitting to historically dominant features and failing to adapt to rapid interest shifts.  Second, a critical information chasm emerges from the point-wise ranking paradigm. By scoring each candidate in isolation, CTR models discard the rich contextual signal implied by the recalled set as a whole, leading to a misalignment where long-term preferences often override the user's immediate, evolving intent.

To address these issues, we propose \textbf{GenCI}, a generative user intent framework that leverages semantic interest cohorts to model dynamic user preferences for CTR prediction. The framework first employs a generative model, trained with a next-item prediction (NTP) objective, to proactively produce candidate interest cohorts. These cohorts serve as explicit, candidate-agnostic representations of a user's immediate intent. A hierarchical candidate-aware network then injects this rich contextual signal into the ranking stage, refining them with cross-attention to align with both user history and the target item. The entire model is trained end-to-end, creating a more aligned and effective CTR prediction pipeline. Extensive experiments on three widely used datasets demonstrate the effectiveness of our approach.

\end{abstract}

\begin{CCSXML}
<ccs2012>
   <concept>
       <concept_id>10002951.10003317.10003347.10003350</concept_id>
       <concept_desc>Information systems~Recommender systems</concept_desc>
       <concept_significance>500</concept_significance>
       </concept>
 </ccs2012>
\end{CCSXML}

\ccsdesc[500]{Information systems~Recommender systems}

\keywords{CTR Prediction, User Interest Modeling, Next-Item Prediction, Recommender Systems }

\maketitle

\newcommand\webconfavailabilityurl{https://doi.org/10.1145/3774904.3792612}
\ifdefempty{\webconfavailabilityurl}{}{
\begingroup\small\noindent\raggedright\textbf{Resource Availability:}\\
% please change the following context to include multiple artifacts if necessary, including data, models, code, etc.
The source code of this paper has been made publicly available at \url{https://doi.org/10.5281/zenodo.18344901}.
\endgroup
}

\input{sec-intro}
\input{sec-method}

\input{sec-experiments}

\input{sec-related}

\section{Conclusion}
In this paper, we present \textbf{GenCI}, a generative user intent framework that leverages semantic interest cohorts to model dynamic user preferences for CTR prediction.  Our framework employs a Transformer-based generative model for the NTP task to produce candidate interest cohorts that embody multi-perspective user intent. A hierarchical candidate-aware network then refines these cohorts via dynamic attention, creating personalized and contextually relevant representations. These refined cohorts are then injected as dynamic contextual signals into the ranking stage, enabling the CTR model to incorporate recall-stage information. The entire model is trained end-to-end under a joint optimization scheme to enhance recall-ranking consistency, and augmented with self-supervised regularization to stabilize learning and improve robustness. Extensive experiments on three benchmark datasets show that GenCI consistently outperforms state-of-the-art baselines, demonstrating its effectiveness at capturing dynamic user intent and improving CTR prediction. For future work, we plan to explore integrating multi-modal and cross-domain information to further enrich semantic ID representations and enhance interest modeling.

\begin{acks}
This paper was partially supported by the National Natural Science Foundation of China No. 92470205 and Beijing Major Science and Technology Project under Contract no. Z251100008425002. Xin Zhao is the corresponding author.
\end{acks}

\bibliographystyle{ACM-Reference-Format}
\balance
\bibliography{main}

\input{sec-appendix}

\end{document}

%% file: sec-intro.tex
\section{Introduction}

The click-through rate (CTR) prediction task, which aims to estimate the probability of a user clicking an item, is essential for web applications like recommender systems to improve user experience and platform revenue. Various models~\cite{zhou2018deep,pi2019practice,zhou2019deep} have been proposed to capture user interests from historical behaviors. Despite the notable achievements, two fundamental challenges persist in contemporary paradigms.

\begin{figure}[t]
  \centering
 \includegraphics[width=0.9\linewidth]{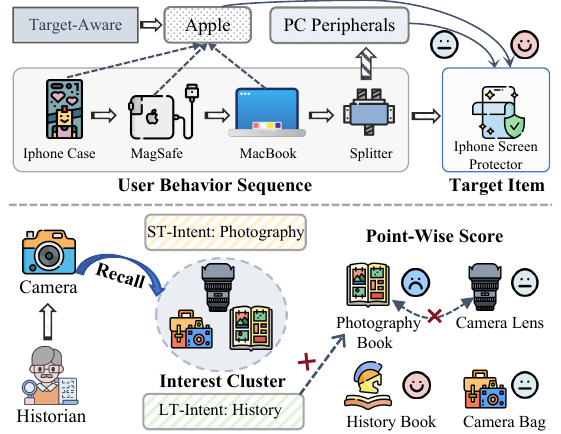}
  \captionsetup{font={small}}
  \caption{Illustration of two key challenges in CTR prediction. Top: Relevance-based models focus on long-term preference matching and fail to adapt to short-term intent shifts (\eg from \textit{iPhone accessories} to \textit{MacBook accessories},
). Bottom: Inconsistency between recall and ranking, where the recall stage retrieves intent-relevant items (\eg photography-related), but the ranking stage favors items aligned with long-term interests (\eg history books).}
  \label{fig:intro}
\end{figure}

First, existing models have limited ability to capture shifts in user interest. Prior studies~\cite{zhou2018atrank,zhou2018deep,zhou2019deep} demonstrate that leveraging behavioral histories can improve performance, primarily by using attention mechanisms to weigh the relevance of past behaviors with respect to the target item. However, these models are designed to assess candidate relevance rather than construct a standalone representation of the user’s current intent, making it difficult to handle evolving interests. As illustrated in the upper part of Figure~\ref{fig:intro}, when a user's focus shifts from \textit{iPhone accessories} to \textit{MacBook accessories}, such models tend to assign a high score to an iPhone screen protector due to the abundance of Apple-related items in the user's history, thereby failing to capture the evolving preference. Moreover, the fundamentally discriminative nature of these approaches often induces shortcut learning~\cite{xu2022alleviating,guo2023embedding,zhang2025dgenctr}, where the model relies on local strongly predictive historical features instead of capturing the full spectrum of user interests, further hindering its ability to model dynamic user preferences.

Second, the point-wise CTR prediction paradigm evaluates each recalled candidate in isolation, ignoring the latent interest signal implied by the candidate set as a whole. In modern \textit{recall-then-rank} pipelines, this context-agnostic approach introduces a substantial information gap. The recall module retrieves a candidate set that reflects hypotheses about the user’s immediate intent, forming an contextual interest cohort(\ie a group of semantically similar items).
However, this rich contextual signal is later discarded under the point-wise ranking schema, leading to misalignment between the two stages. As illustrated in the lower part of Figure~\ref{fig:intro}, when a user shifts their focus from history to photography, relevant photography books may be successfully recalled, yet the context-agnostic ranker might still prioritize familiar history items driven by strong long-term preferences, failing to capture the user’s evolving intent.

To overcome these limitations, we introduce a generative paradigm to more effectively model user interest shifts through an auxiliary next-item prediction (NTP) task. Our approach is grounded in modeling the user’s chronologically ordered click sequence, through which the causal decision-making process and evolving intent can be inferred~\cite{liu2024at4ctr}. Based on this foundation, our model proactively generates semantic IDs that represent a hypothesis about the user's next interaction, overcoming the limitation of discriminative methods that passively evaluate target relevance without forming a candidate-agnostic representation of current intent.   Furthermore, because this generative prediction is made over the entire item space~\cite{rajput2023recommender}, the resulting semantic IDs encode globally coherent information about user interests, mitigating the overfitting of discriminative methods to local high-frequency features. By modeling the interaction between these semantic IDs and the target item, we inject this rich contextual foundation into the CTR prediction stage, enabling the ranking model to be aware of the recall-stage context (\ie why an item was retrieved alongside others). This design addresses the inherent limitation of point-wise CTR models, which evaluate items independently without awareness of the retrieval context, thereby achieving better alignment between the recall and ranking stages.

% This design breaks the isolation of traditional point-wise CTR prediction, achieving stronger consistency and alignment between recall and ranking.  

In this paper, we propose \textbf{GenCI}, a novel generative user intent framework that constructs semantic interest cohorts as explicit intent representations for CTR prediction.
First, hierarchical quantization is applied to organize items into semantically coherent cohorts, forming latent interest groups that capture fine-grained user preferences. Based on this representation, we employ a generative sequential model built upon a Transformer architecture to perform the NTP task, which extracts dual-aspect user intent: the encoder produces a general long-term interest profile, while the decoder generates candidate interest cohorts that capture the user’s immediate, short-term intent. These candidate cohorts are then incorporated as dynamic contextual signals in the ranking stage. To mitigate the noise introduced by items selected merely due to coarse semantic similarity, we design a hierarchical candidate-aware modeling module that refines the cohorts' intent via cross-attention, aligning them with both historical behaviors and the target item to produce highly personalized and dynamic interest representations. Finally, the multi-perspective intent signals are integrated with conventional features and fed into MLPs for CTR estimation, with the entire model trained end-to-end under a joint optimization scheme, augmented by a self-supervised regularization that stabilizes learning and improves robustness.
The contributions of this paper are summarized as follows:

$\bullet$ We propose \textbf{GenCI}, a novel framework that introduces a generative paradigm for modeling dynamic user intent. It leverages a NTP task to proactively generate semantically coherent interest cohorts, enabling the capture of multi-faceted user preferences.

$\bullet$ We introduce a hierarchical candidate-aware network that personalizes the generated cohorts via cross-attention, thereby injecting them as dynamic signals to make the point-wise ranker aware of the recall signal.

$\bullet$ We conduct extensive experiments on three widely used datasets. GenCI consistently outperforms a number of competitive baselines, showing the effectiveness of our model.

%% file: sec-method.tex
\begin{figure*}[ht]
  \centering
  \includegraphics[width=0.95\textwidth]{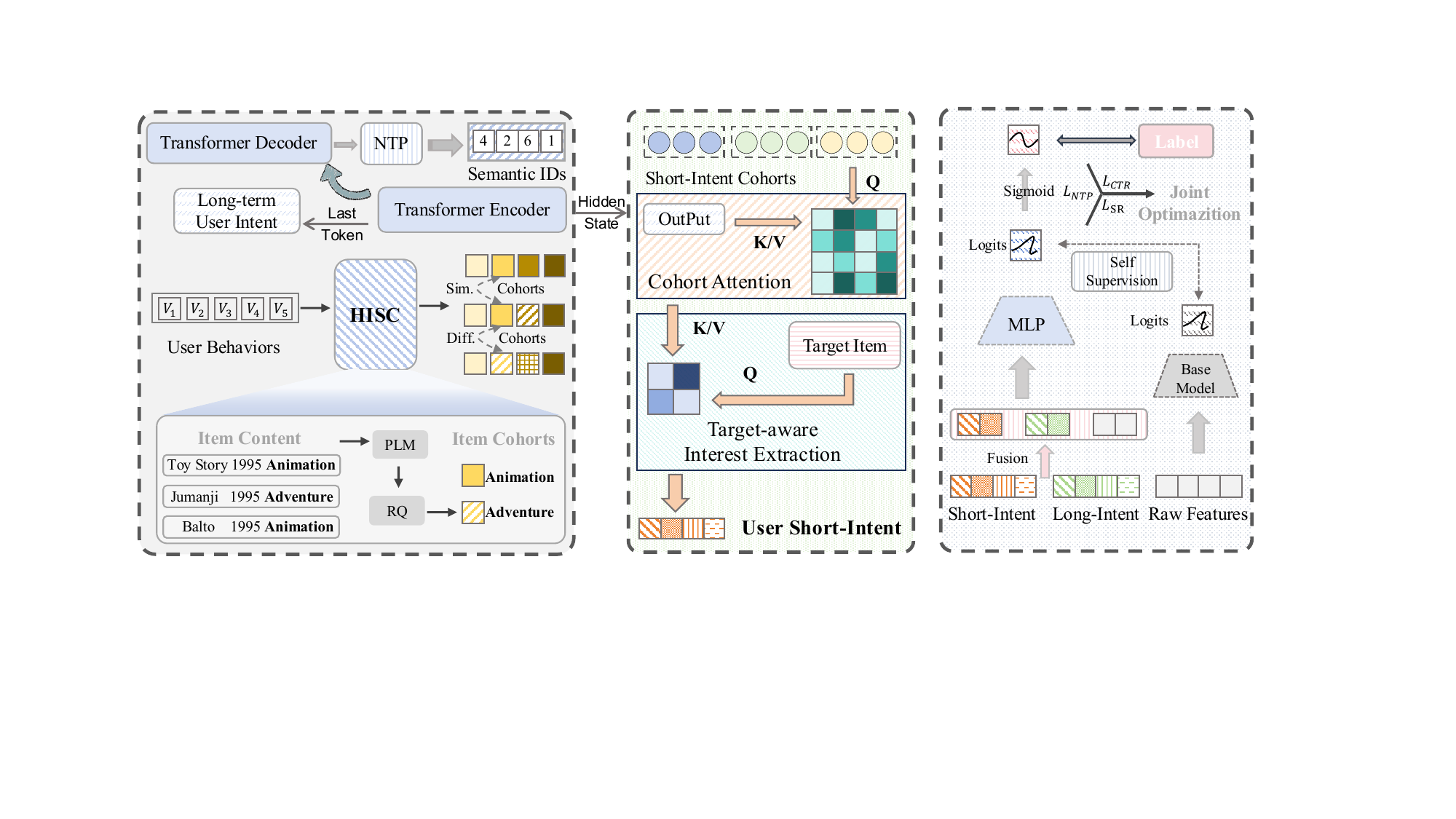}
  \captionsetup{font={small}}

\caption{Overview of the proposed \textbf{GenCI} framework. The left part illustrates the generative module that constructs user interest cohorts, where ``HISC'' generates these cohorts via hierarchical quantization. The middle part details the Hierarchical Candidate-Aware Intent Modeling, which refines these cohorts. The right part showcases multi-intent fusion and jointly optimized with self-supervision for CTR prediction.}
  
  % \caption{Overview of the proposed \textbf{GenCI} framework. \textbf{Hierarchical Interest Semantic Cohorts (HISC)} group items into coherent semantics via RQ-VAE. Candidate cohorts representing short-term intents are refined through hierarchical attention and jointly optimized with self-supervision for CTR prediction.} 
  \label{fig:framework}
\end{figure*}

\section{Methodology}
In this section, we present the proposed approach, named \textbf{GenCI} (Figure~\ref{fig:framework}), which adopts a generative paradigm for multi-faceted user intent modeling. Our framework first employs a generative module to produce candidate interest cohorts that capture the user’s short-term intent. Building on this, a hierarchical interest network is applied to extract and refine the most behavior-relevant and target-aligned signals for the final recommendation.  This entire process, stabilized by a self-supervised regularization objective, is designed to achieve end-to-end consistency between recall and ranking for CTR prediction. Subsequently, we first introduce the task
formulation (Section~\ref{sec:task}), followed by in-depth discussions on  generative user-intent modeling (Section~\ref{Generate User-Itent}) and interest extraction and utilization (Section~\ref{Interest Extraction}).

\subsection{Task Formulation}
\label{sec:task}
\paratitle{CTR Predition Task.}
\label{sec:ctr task}
Click-through rate~(CTR) prediction aims to predict the probability that a user clicks on an item.
Typically, it is formulated as a binary classification task, in which the click label (\ie \emph{Yes} or \emph{No}) is to be predicted. 
Formally, each instance is represented as an input–label pair $\langle \bm{x}, y \rangle$, where the input $\bm{x}$ consists of three types of features: user attributes $\bm{x}_u$ (\eg user ID, gender), item profiles $\bm{x}_i$ (\eg item ID, category), and interaction context $\bm{x}_c$ (\eg device type, timestamp). 
The label $y \in \{0,1\}$ indicates whether a click actually occurred. 
Given a CTR prediction model $f(\cdot)$, it takes as input all feature vectors $\bm{x} = \{\bm{x}_u, \bm{x}_i, \bm{x}_c\}$ and outputs the predicted click probability $\hat{y} = f(\bm{x})$.
To optimize the prediction model $f(\cdot)$, we can minimize  the 
binary cross-entropy~(BCE) loss between the ground-truth label and the predicted label over all the training instances, which is defined as follows:  
 \begin{equation}
  \label{equ:ctr}
    \mathcal{L}_{CTR} = -\frac{1}{N} \sum_{i=1}^{N}\bigg(y_i\log(\hat{y}_i)+(1-y_i)\log(1-\hat{y}_i)\bigg). 
  \end{equation}

\paratitle{Next-Item Predition Task.}
\label{sec:ntp task}
Given an item set $\mathcal{V}$, a user’s chronological interaction history is represented as $S = [v_1, v_2, \dots, v_t]$. Sequential recommendation aims to infer the user’s implicit preferences from this sequence and predict the next item of interest, $v_{t+1}$.
% We refer to this formulation as the NTP task. 
Generative recommendation reformulates this task as a sequence-to-sequence problem.
Specifically, an item tokenizer $T$ is trained to map each item $v \in \mathcal{V}$ to a token-based identifier, a process known as item tokenization. Formally, this mapping is denoted as $T(v) = [c_1, \dots, c_N]$, where each $c_n$ is the $n$-th token in the identifier and $N$ is the identifier length.
Then, the historical sequence $S$ and the target item $v_{t+1}$ are tokenized into input and output sequences $X = T(S) = [c^1_1, c^1_2, \dots, c^t_N]$ and $Y = T(v_{t+1}) = [c^{t+1}_1, \dots, c^{t+1}_N]$, respectively. The prediction task is then performed by autoregressively generating the token sequence of the target item. This process is formally expressed as:
\begin{equation}
P(Y \mid X) = \prod_{n=1}^{N} P\left(c^{t+1}_n \mid X, c^{t+1}_1, \dots, c^{t+1}_{n-1}\right).
\end{equation}
In practice, this objective is optimized using cross-entropy loss, denoted as $L_{NTP}$.

\subsection{Generative User-Intent Modeling}
\label{Generate User-Itent}

Previous CTR prediction models primarily employ a discriminative paradigm focused on relevance matching~\cite{zhou2018deep}. This makes them susceptible to shortcut learning~\cite{xu2022alleviating,guo2023embedding}, where they overfit to historically dominant features and  fail to adapt to sudden interest shifts. Inspired by the principle that past behaviors sequentially influence current actions~\cite{liu2024at4ctr}, we propose a generative approach using the NTP task to model these transitional patterns. By predicting over the entire item pool, the model learns a globally coherent semantic structure, going beyond the local pattern-matching of discriminative methods. From these globally-aware semantic ID representations, we construct user interest cohorts to capture each user’s immediate and contextually relevant intent effectively.

\subsubsection{Semantic Interest Cohorts via Hierarchical Quantization}

To organize the item space into semantically coherent groups, we adopt the residual-quantized variational autoencoder (RQ-VAE)~\cite{zeghidour2021soundstream} as the item tokenizer, following the design in TIGER~\cite{rajput2023recommender}.
This approach performs multi-level residual quantization on textual embeddings derived from item titles and descriptions, producing hierarchical discrete representations known as semantic IDs.
The tokenizer consists of $N$ hierarchical quantization levels indexed by $n \in \{1,\dots,N\}$.
Each level is associated with an independent codebook $\mathcal{C}^n = \{\bm{c}_k^n\}_{k=1}^{K}$, where $\bm{c}_k^n$ denotes the $k$-th codeword at the $n$-th quantization level, and $K$ is the codebook size.
At the $n$-th level, the residual vector $\bm{r}^{(n)}$ is quantized by assigning it to the nearest codeword:
\begin{align}
    & z_n = \underset {k} { \operatorname {arg\,min} } ||\bm{r}^{(n)} - \bm{c}_k^n||_2^2
, 
 \quad \bm{r}^{(n+1)} = \bm{r}^{(n)} - \bm{c}_{z_n}^n,
    \label{eq:rq}
\end{align}
where $z_n$ is the selected codeword index at level $n$.
Crucially, all items assigned the same codeword at level $n$ (\ie $z_n = k$) form a semantic \textbf{cohort}, capturing shared semantic characteristics at that granularity.
Across all levels, the hierarchical quantization induces $N \times K$ semantic cohorts, organizing the item corpus into multi-granularity interest groups that serve as the foundation for subsequent short-term intent modeling.

\subsubsection{Dual-Aspect Interest Extraction from Behavioral Sequences.}
\label{sec:interest cohorts}
Leveraging the semantic ID representations, we employ a Transformer-based encoder to model user behavioral sequences for generative retrieval. The encoder, composed of $L$ layers, processes the input sequence $\bm{x}$ to produce a stack of hierarchical representations $\mathbf{H}_{\text{enc}}$:
\begin{equation}
  \label{equ:encoder}
  \text{Encoder}(\bm{x}) = [\mathbf{H}^{(1)}, \mathbf{H}^{(2)}, \dots, \mathbf{H}^{(L)}],
\end{equation}
where $\mathbf{H}^{(l)}$ denotes the hidden states at layer $l$. The final layer's output $\mathbf{H}^{(L)}$ encapsulates the deepest contextual information. To obtain the user's long-term intent ($\text{LTI}$) representation $\mathbf{H}_{\text{LTI}}$, we extract the hidden state from $\mathbf{H}^{(L)}$ at the last valid position:
% \begin{equation}
% \label{equ:enecoder_last}
%     \mathbf{H}_{\text{LTI}} = \mathbf{h}^{(L)}_{P}, \quad \text{where } P = \sum_{i=1}^{N} m_i,
% \end{equation}
\begin{equation}
\label{equ:enecoder_last}
    \mathbf{H}_{\text{LTI}} = \bm{h}^{(L)}_{P}, \quad \text{where } P = \max \{i \mid m_i = 1\},
\end{equation}
where $P$ denotes the last valid position determined by the mask $\bm{m}$.

During training, the model captures users’ short-term and evolving interests by predicting the semantic IDs of the next item in the NTP task.  Specifically, the Transformer decoder conditions on the encoder output $\mathbf{H}_{\text{enc}}$ and the embedded target prefix  $\mathbf{E}_{{<t}}$ at step $t$ to generate the current preference state: 
\begin{equation}
\label{decoder}
\mathbf{H}_{\text{dec}}^{(t)} = \text{Decoder}_{\mathcal{R}}(\mathbf{H}_{\text{enc}}, \mathbf{E}_{{<t}}),
\end{equation}
which is subsequently projected onto the codebook space to compute the probability distribution over the next semantic ID. Leveraging this learned sequential knowledge, the model can autoregressively generate a sequence of valid semantic IDs for a given user’s historical interactions, using a constrained beam search to ensure that the generated IDs correspond to items within the recommendation corpus.

This approach capitalizes on the Transformer’s ability to capture short-term dependencies~\cite{li2023gpt4rec}. By formulating the prediction target as structured semantic IDs rather than atomic item identifiers, the decoder is encouraged to learn abstract user intents rather than superficial item co-occurrence patterns.
Furthermore, the autoregressive attention mechanism naturally prioritizes recent interactions, ensuring that the generated cohorts accurately reflect users’ immediate interests. Consequently, the NTP task effectively models sequential behavior dynamics, capturing the user's evolving short-term intent, complementing the static long-term profile $\mathbf{H}_{\text{LTI}}$.

\subsection{Interest Extraction and Utilization}
\label{Interest Extraction}

To fully exploit the contextual information captured by the squential modeling module, we treat the above generated candidate set as a dynamic contextual signal to guide CTR prediction. In this part, 
We propose a \textit{hierarchical candidate-aware interest modeling} module that explicitly integrates this short-term intent into the CTR prediction process, improving alignment between recall and ranking.  Subsequently, we aggregate multiple signals, including long-term preferences, short-term intent, and conventional raw features (e.g., user profile and target item), feeding them into MLPs to model high-order feature interactions. Finally, the entire framework is trained jointly with both generative and predictive objectives.

\subsubsection{Hierarchical Candidate-Aware Intent Modeling}

A critical challenge in leveraging the generated candidate set is that it may include items selected merely due to coarse semantic similarity, rather than genuine relevance to the user.Consequently, naïve aggregation can introduce irrelevant signals. To address this, we employ a hierarchical attention mechanism to distill a refined intent representation that is tightly aligned with both the user’s historical context and the specific target item under prediction.

\paratitle{Candidate Cohort-Guided Intent Refinement.}
Building upon the candidate generation in Section~\ref{sec:interest cohorts}, which reflects the user’s immediate preferences, we further distill the underlying short-term intent through a behavior-aware refinement process. This step is crucial for grounding the semantically derived interest cohorts in the user's concrete behavioral patterns.
Specifically, the generated sequence of semantic IDs is mapped to hierarchical item cohorts $\{C_1, C_2, \dots, C_{N-1}\}$. Here, each $C_n$ denotes the set of items sharing the predicted codeword at level $n$, thereby capturing a specific granularity of the user's short-term intent. To preserve this hierarchical structure, we apply mean pooling individually within each cohort to derive the \textit{cohort-level representation} $\mathbf{E}_{\text{C}}$:
\begin{equation}
  \label{equ:cohort_mean}
  \bm{e}_{\text{C}}^{(n)} = \text{MeanPool}(\{\bm{e}_i \mid i \in C_n\}), \quad \mathbf{E}_{\text{C}} = [\bm{e}_{\text{C}}^{(1)}; \dots; \bm{e}_{\text{C}}^{(N-1)}],
\end{equation}
where $\mathbf{E}_{\text{C}} \in \mathbb{R}^{(N-1) \times d}$ represents the multi-level cohort embeddings.
To incorporate behavior-relevant context, we compute the \textit{refined intent representation}, denoted as $\mathbf{E}_{\text{R}}$, by employing the cohort matrix $\mathbf{E}_{\text{C}}$ as a query to attend to the user's historical context $\mathbf{H}_{\text{enc}}$:
\begin{equation}
\label{equ:cross_attn_v1}
\mathbf{E}_{\text{R}} =
\operatorname{Softmax}\left( \frac{(\mathbf{E}_{\text{C}} \mathbf{W}_Q)(\mathbf{H}_{\text{enc}} \mathbf{W}_K)^{\top}}{\sqrt{d}} \right) (\mathbf{H}_{\text{enc}} \mathbf{W}_V),
\end{equation}
where $\mathbf{W}_{\{Q,K,V\}}$ are learnable projection matrices and subscripts $\text{C}$ and $\text{R}$ stand for \textit{cohort} and \textit{refined}, respectively.
Mechanistically, this operation re-weighs the user's historical actions based on their relevance to the predicted cohort. This selectively emphasizes historical signals that support the current semantic hypothesis while suppressing behaviorally irrelevant noise. The resulting $\mathbf{E}_{\text{R}}$ thus encodes a behaviorally grounded short-term intent, providing a highly discriminative signal for the subsequent ranking stage.

\paratitle{Target-Aware Dynamic Interest Extraction.}  
\label{sec:TAD interset}
Although the previous step filters out many irrelevant signals (\ie items included in the cohorts merely due to semantic similarity), the resulting multi-level representation $\mathbf{E}_{\text{R}}$ preserves a hierarchical structure that requires adaptive aggregation. A static strategy, such as mean pooling across the $N$ hierarchical levels, would yield a target-agnostic interest vector. This is suboptimal, as the importance of different semantic granularities naturally varies depending on the specific target item (e.g., broad categories vs. specific niches). To address this limitation, we introduce a target-aware attention mechanism. We utilize the ID embedding $\bm{e}_{v}$ of the target item $v$ as a query to attend to the refined cohort sequence $\mathbf{E}_{\text{R}}$. Formally, the dynamic interest representation is computed as:  
\begin{equation}
\label{equ:dyna}
\mathbf{H}_{\text{STI}} =
\operatorname{Softmax}\left( \frac{(\bm{e}_{v} \mathbf{W}_Q)(\mathbf{E}_\text{R} \mathbf{W}_K)^{\top}}{\sqrt{d}} \right) (\mathbf{E}_\text{R} \mathbf{W}_V),
\end{equation}
 where $\mathbf{W}_{\{Q,K,V\}}$ are learnable projection matrices. Through this mechanism, the model adaptively highlights the specific cohort levels most relevant to the target, yielding a precise and context-aware short-term interest ($\text{STI}$) representation $\mathbf{H}_{\text{STI}}$ for the final prediction.

\subsubsection{Multi-Intent Fusion and Joint Optimization.}
Building on the multi-perspective user intent representations captured in previous stages, this part integrates them with conventional features for CTR prediction and further introduces a self-supervised regularization to stabilize learning. Finally, the entire model is trained end-to-end via a joint optimization scheme, which facilitates a seamless information exchange across different modules and ensures coherent alignment between intent modeling and prediction objectives.

\paratitle{Multi-faceted Interest Aggregation for Prediction.}
In the final prediction layer, we aggregate the multi-faceted interest representations derived from the preceding modules. Specifically, we concatenate the long-term intent $\mathbf{H}_{\text{LTI}}$ (derived in Eq.\eqref{equ:enecoder_last}) and the target-aware short-term intent $\mathbf{H}_{\text{STI}}$ (computed in Eq.\eqref{equ:dyna}) with conventional feature embeddings to estimate the final CTR score:
\begin{equation}
\label{equ:pred_v1}
\hat{y} = \sigma \left( \operatorname{MLP}\left( [\mathbf{H}_{\text{LTI}}; \mathbf{H}_{\text{STI}}; \mathbf{E}_{u}; \mathbf{E}_{v}] \right) \right),
\end{equation}
where $\mathbf{E}_{u}$ and $\mathbf{E}_{v}$ denote the concatenated feature embeddings for the user $u$ and the target item $v$, respectively. The operator $[\cdot ; \cdot]$ represents concatenation, $\operatorname{MLP}(\cdot)$ is the prediction network, and $\sigma(\cdot)$ is the sigmoid function mapping the output to a click probability.

\paratitle{Joint Optimization.} 
Building upon the multi-faceted interest representations introduced above, we further optimize the entire model through a self-supervised joint objective. To enhance training stability, particularly in the early stages when the sequence modeling module may produce noisy intent representations, we incorporate an auxiliary regularization term. Specifically, a lightweight baseline CTR model (e.g., DeepFM) trained on raw features provides supervisory logits, which are used to regularize the full model’s output and guide the integrated intent representations toward more reliable patterns:
 \begin{equation}
  \label{equ:sdr}
 \mathcal{L}_{\text{SR}} = -\frac{1}{N} \sum_{i=1}^{N}
 L\bigg(
\textsf{OP}\big(\mathcal{M}_{\text{base}}(\bm{x}_i)\big),
\textsf{OP}\big(\mathcal{M}_{\text{ours}}(\bm{x}_i, \boldsymbol{\phi}_i)\big)
 \bigg),
\end{equation}
where $\mathcal{M}_{\text{base}}$ and $\mathcal{M}_{\text{ours}}$ denote the baseline and proposed CTR models, respectively, $\textsf{OP}(\cdot)$ extracts logits, and $\boldsymbol{\phi}_i$ 
denotes the multi-level intent representation(\ie $\mathbf{H}_{\text{LTI}}$ and $\mathbf{H}_{\text{STI}}$) of instance $i$.
Finally, the framework is optimized in an end-to-end manner under a multi-task learning paradigm, incorporating the primary CTR prediction task, the NTP task(defined in Section \ref{sec:task}), and the self-supervised regularization (SR) task, with the overall loss defined as:
 \begin{equation}
  \label{equ:joint}
\mathcal{L} = \mathcal{L}_{\text{CTR}} + \mu \mathcal{L}_{\text{NTP}} +  \eta \mathcal{L}_{\text{SR}},
\end{equation}
where $\mu$ and $\eta$ are hyperparameters that balance the contributions of each loss term. This unified optimization allows gradients to propagate across the entire architecture, fostering a synergistic feedback loop. Consequently, the model jointly refines long-term and short-term user interests while aligning the recall-oriented and ranking-oriented objectives under a single global objective, leading to more robust end-to-end CTR prediction.

\subsection{Discussion}
In this section, we discuss the distinct advantages and architectural insights of the proposed GenCI framework.

\paratitle{Recall-Ranking Consistency.}
Conventional CTR prediction systems suffer from a key limitation originating from the ranker's point-wise scoring paradigm. Because the ranker scores each candidate independently, it loses the rich contextual signal that is inherent in the co-recalled items. Our framework overcome this limitation by reframing the pipeline from  \textit{recall-then-rank} process into an integrated \textit{generate-and-interpret} loop. Our generative module produces a semantically coherent interest cohort to act as a dynamic, contextual prior for the ranker. The entire architecture is jointly optimized, which helps the recall stage generate cohorts that are most informative for the ranking stage, ultimately achieving effective recall-ranking consistency.

\paratitle{Cohort-Level Intent Modeling.}
Unlike conventional methods that infer user interest solely from individual historical items, our framework introduces a cohort-level intent modeling paradigm. We exploit a key property of generative sequence modeling: Semantic IDs naturally group semantically similar items through shared codes. By mapping generated Semantic IDs back to their corresponding items, we construct interest cohorts that provide a broader reflection of the user’s immediate intent. To refine these cohorts, we employ a hierarchical attention mechanism to suppress items that are semantically correlated but lack behavioral evidence, resulting in user representations that are both behaviorally grounded and target-aware.

The overall comparison of our approach with existing methods
is presented in Table ~\ref{tab:discussion}.

\begin{table}[!h]
\centering
    \captionsetup{font={small}}
    \caption{Comparison of different CTR methods. ``ST'', ``LT'' refer short-term and long-term respectively, CL represents cluster-level, and ``RRC'' denotes recall–ranking consistency}
    \label{tab:discussion}
    \resizebox{0.85\columnwidth}{!}{
    \begin{tabular}{l|ccccc}
    \toprule
    Methods	& LT Intent & ST Intent & CL Intent & RRC   &  \\
		\hline
     DCNV2~\cite{wang2021dcn}  & \textcolor{teal}{\CheckmarkBold} &	\textcolor{purple}{\XSolidBrush} &	 \textcolor{purple}{\XSolidBrush} & \textcolor{purple}{\XSolidBrush} \\  
     DIN~\cite{zhou2018deep}  & \textcolor{teal}{\CheckmarkBold} &	\textcolor{purple}{\XSolidBrush} &	 \textcolor{purple}{\XSolidBrush} & \textcolor{purple}{\XSolidBrush} \\
      DIEN~\cite{zhou2019deep}  & \textcolor{teal}{\CheckmarkBold} &	 \textcolor{teal}{\CheckmarkBold} &	 \textcolor{purple}{\XSolidBrush} & \textcolor{purple}{\XSolidBrush} \\ 
            % & \textcolor{purple}{\XSolidBrush} &	\textcolor{teal}{\CheckmarkBold} &	 \textcolor{purple}{\XSolidBrush} & \textcolor{purple}{\XSolidBrush} \\

    GenCI~(ours)	&	\textcolor{teal}{\CheckmarkBold} &	\textcolor{teal}{\CheckmarkBold} &	\textcolor{teal}{\CheckmarkBold} & \textcolor{teal}{\CheckmarkBold}  \\
\bottomrule
\end{tabular}
}
\vspace{-1em}
\end{table}

%% file: sec-experiments.tex
\section{Experiments}

In this section, we conduct extensive experiments to evaluate the effectiveness of \textbf{GenCI} and analyze the impact of each model component.

\subsection{Experimental Setup}
We introduce the experimental settings, including the datasets,
baseline approaches, metrics, and hyperparameter details.

\subsubsection{Datasets}
We conduct experiments on three public datasets for recommender evaluation, including: movie recommendation (MovieLens\footnote{https://grouplens.org/datasets/movielens}) and product recommendation (two subsets from Amazon\footnote{https://amazon-reviews-2023.github.io/}, namely Fashion and Musical-Instruments). 
The statistics of datasets are summarized in Table~\ref{tab:datasets}.

$\bullet$ \textbf{MovieLens} is a widely used benchmark containing users’ movie rating records. Following prior work~\cite{tian2024rotative,cheng2020adaptive}, ratings above 3 are regarded as positive interactions and those below 3 as negative. User interactions are time-ordered and split into training, validation, and test sets (8:1:1), with each sequence truncated to 50 items.\par
$\bullet$ \textbf{Fashion} and \textbf{Instrument} contain user reviews from May 1996 to September 2023. Following~\cite{zhou2018deep}, we filter users and items with fewer than five interactions, then apply a leave-one-out strategy with each user's sequence limited to 20 items.

\begin{table}[h]
  \captionsetup{font={small}}
  \caption{The statistics of datasets. 
  The``Avg.\textit{n}'' is the
average length of behavioral sequences.} 
  \label{tab:datasets}
  \resizebox{.45\textwidth}{!}{
  \begin{tabular}{c|ccccc}
    \toprule
    \textbf{Datasets}& $\#$ Users & $\#$ Items & $\#$ Interaction  &  Avg.\textit{n} &Sparsity \\
    \hline \hline
    MovieLens & 6041 & 3669 & 739012 &122.35 &96.66\% \\
    Fashion & 3719 & 13198 &24097 & 6.48 & 99.95\%  \\
    Instrument & 57440  & 24588 &511836 & 8.91 & 99.96\% \\
  \bottomrule
\end{tabular}}
\end{table}

\begin{table*}[t]
\centering
  \small
  \captionsetup{font={small}}
\caption{Performance comparisons. Note that a higher AUC or lower LogLoss at 0.001-level is significant for CTR prediction.}
% ~\cite{chen2021enhancing,huang2019fibinet,luo2020network,lian2018xdeepfm,wang2020dcn}.}
\label{table:performance}
\resizebox{0.9\textwidth}{!}{
\begin{tabular}{c|cccc|cccc|cccc}
\toprule
\multirow{2}{*}{\textbf{Model}} &
\multicolumn{4}{c|}{\textbf{MovieLens}} &
\multicolumn{4}{c|}{\textbf{Amazon-Fashion}} &
\multicolumn{4}{c}{\textbf{Amazon-Instrument}}\\
& AUC & LogLoss & RelaImpr & Latency & AUC & LogLoss & RelaImpr & Latency & AUC & LogLoss & RelaImpr & Latency\\
\hline\hline
LR          & 0.8577 & 0.3904 & --     & 0.21s & 0.7284 & 0.5097 & --      & 0.01s & 0.6985 & 0.4200 & --     & 0.16s \\
FwFM        & 0.8743 & 0.3694 & 4.64\% & 0.51s & 0.7317 & 0.5947 & 1.44\%  & 0.01s & 0.7118 & 0.4075 & 6.70\% & 0.19s \\  
NFM         & 0.8847 & 0.3649 & 7.55\% & 0.43s & 0.7421 & 0.5239 & 6.00\%  & 0.03s & 0.7022 & 0.3937 & 1.86\% & 0.25s \\
PNN         & 0.8901 & 0.3511 & 9.06\% & 0.31s & 0.7440 & 0.5842 & 6.83\%  & 0.02s & 0.7144 & 0.3875 & 8.01\% & 0.17s \\
FiGNN       & 0.8899 & 0.3628 & 9.00\% & 0.45s & 0.7445 & 0.6632 & 7.05\%  & 0.02s & 0.7146 & 0.3958 & 8.11\% & 0.30s \\
DeepFM      & 0.8908 & 0.3527 & 9.25\% & 0.41s & 0.7430 & 0.6949 & 6.39\%  & 0.02s & 0.7151 & 0.3866 & 8.36\% & 0.22s \\
DCNV2       & 0.8927 & 0.3564 & 9.78\% & 0.38s & 0.7423 & 0.6147 & 6.08\%  & 0.02s & 0.7148 & 0.3869 & 8.21\% & 0.21s \\
AutoInt+    & 0.8924 & 0.3534 & 9.70\% & 0.62s & 0.7446 & 0.6218 & 7.09\%  & 0.05s & 0.7157 & 0.3862 & 8.66\% & 0.26s \\
xDeepFM     & 0.8919 & 0.3512 & 9.56\% & 0.47s & 0.7443 & 0.6465 & 6.96\%  & 0.03s & 0.7158 & 0.3868 & 8.71\% & 0.31s \\
AFN+        & 0.8906 & 0.3574 & 9.19\% & 0.58s & 0.7439 & 0.5559 & 6.79\%  & 0.02s & 0.7102 & 0.3899 & 5.89\% & 0.23s \\
% Final       & 0.8917 & 0.3549 & 9.50\% & 0.46s & 0.7447 & 0.5875 & 7.13\%  & 0.03s & 0.7161 & 0.4128 & 8.86\% & 0.28s \\
RFM         & 0.8918 & 0.3612 & 9.53\% & 0.44s & 0.7463 & 0.4903 & 7.84\%  & 0.02s & 0.7131 & 0.3879 & 7.35\% & 0.23s \\
DIN         & 0.8931 & 0.3445 & 9.89\% & 0.52s & 0.7445 & 0.7838 & 7.05\%  & 0.03s & 0.7166 & 0.3894 & 9.11\% & 0.29s \\
DIEN        & 0.8936 & 0.3459 & 10.03\% & 2.34s & 0.7451 & 0.8129 & 7.31\%  & 0.09s & 0.7168 & 0.3879 & 9.21\% & 1.06s \\
MIRRN & 0.8938 & 0.3472 & 10.05\%  & 0.65s & \underline{0.7466} & 0.5647 & 7.95\% & 0.04s & \underline{0.7169} & 0.3863 & 9.26\%   & 0.36s \\
SFG & \underline{0.8939} & 0.3463 &   10.08\% & 0.50s & 0.7450 & 0.5672 &  7.28\%  & 0.03s & 0.7159 & 0.3920 &  8.79\%  & 0.29s \\

\hline \hline
GenCI & \textbf{0.8964} & 0.3507 & 10.82\% & 0.48s & \textbf{0.7519} & 0.4959 & 10.29\% & 0.03s & \textbf{0.7182} & 0.3849 & 9.92\% & 0.28s \\
\bottomrule
\end{tabular}
}
\end{table*}

\subsubsection{Compared Methods}

We compare the proposed approach with the following baseline methods:

$\bullet$ \textbf{LR}~\cite{richardson2007predicting} captures linear dependencies among features via weighted summation.\par
$\bullet$ \textbf{FwFM}~\cite{sun2021fm2} enhances factorization machines by introducing field-specific parameters.\par
$\bullet$ \textbf{NFM}~\cite{he2017neural}integrates the linearity of FM for second-order feature interactions with the non-linearity of neural networks to capture higher-order interactions effectively.\par
$\bullet$ \textbf{PNN}~\cite{qu2017product} incorporates a product layer to capture interactions between inter-field feature categories.\par
$\bullet$ \textbf{FiGNN}~\cite{li2019fi} models multi-field feature interactions using graph neural networks to capture complex relationships between features.\par
$\bullet$ \textbf{DeepFM}~\cite{guo2017deepfm} unifies FM and MLP to jointly capture explicit and implicit feature interactions.\par
$\bullet$ \textbf{DCNv2}~\cite{wang2021dcn} models high-order interactions via deep cross networks while using MLPs for implicit interactions.\par
$\bullet$ \textbf{AutoInt+}~\cite{song2019autoint} uses multi-head self-attentive neural network to explicitly model the feature interactions. AutoInt+ improves it with a feed-forward neural network.\par
$\bullet$ \textbf{xDeepFM}~\cite{lian2018xdeepfm} employs the Compressed Interaction Network (CIN) to capture vector-wise explicit feature interactions.\par
$\bullet$ \textbf{AFN+}~\cite{cheng2020adaptive} learns arbitrary-order feature interactions in logarithmic space. AFN+ integrates AFN and DNN for better stability.\par
% $\bullet$ \textbf{Final}~\cite{mao2023finalmlp} introduces a factorized interaction layer that unifies feature interaction learning and MLPs within a single block.\par
$\bullet$ \textbf{RFM}~\cite{tian2024rotative} models adaptive-order feature interactions by representing features as polar angles in the complex plane and capturing their relationships through complex rotations.\par
$\bullet$ \textbf{DIN}~\cite{zhou2018deep} employs a local activation unit to adaptively capture user interests from historical behaviors with respect to each target item, enabling instance-level interest representation\par
$\bullet$ \textbf{DIEN}~\cite{zhou2019deep} introduces an auxiliary-supervised interest extractor and an interest evolution layer to capture users’ dynamic and evolving interests more accurately.\par
$\bullet$ 
\textbf{MIRRN}~\cite{xu2025multi} employs multi-granularity queries to retrieve diverse interest subsequences and uses a multi-head Fourier transformer to model long-term user behaviors.\par
$\bullet$ \textbf{SFG}~\cite{yin2025feature} reformulates CTR prediction as a supervised generative feature-generation paradigm, generating feature embeddings conditioned on all features to reduce embedding collapse.

The above baseline methods cover a wide spectrum of approaches in CTR prediction. LR and FwFM capture linear and pairwise interactions, while high-order models like NFM, PNN, and FiGNN model complex non-linear dependencies.
% The above baseline methods cover a wide spectrum of approaches in CTR prediction. LR and FwFM capture linear and pairwise feature interactions, respectively, while high-order interaction models such as NFM, PNN, and FiGNN aim to model more complex, non-linear dependencies among multiple features. 
Ensemble-based models, including DeepFM, DCNv2, xDeepFM, AutoInt+, integrate explicit and implicit interactions for complementarity. AFN and RFM learn arbitrary-order interactions using mathematical transformations, and SFG refines feature embeddings through a generative paradigm.
User behavior models, such as DIN and DIEN, capture evolving interests via target-aware representations, with MIRRN further modeling multi-granularity interests.
% Finally, user behavior models focus on historical dynamics: 
% DIN and DIEN capture evolving interests through target-aware representations, and MIRRN further considers the user’s multi-granularity interests.
% Finally, user behavior models such as DIN and DIEN further capture temporal and evolving user interests, providing target-aware and dynamic representations.

\subsubsection{Evaluation Metrics}
Following previous works~\cite{huang2019fibinet,song2019autoint,sun2021fm2}, we adopt two widely used ranking metrics in CTR models: \textbf{AUC}~(Area Under the ROC curve) and \textbf{LogLoss}~(cross entropy).
The \textbf{RelaImpr} metric, as proposed in~\cite{Yan2014CoupledGL}, measures the relative improvement over the base model and is defined as follows:
\begin{equation}
    \textit{RelaImpr} = \left(\frac{\textit{AUC}(\textit{measure model}) - 0.5}{\textit{AUC}(\textit{base model}) - 0.5} - 1\right) \times 100\%
\end{equation}

\subsubsection{Implementation Details}
We implement all models using the RecBole~\cite{zhao2021recbole} framework. We perform a grid search to find optimal hyperparameters for each model. For sequence modeling, semantic embeddings for items are first learned by encoding their textual information using Sentence-T5~\cite{ni2021sentence}. Based on these embeddings, an RQ-VAE is trained with three codebooks of size 256 and an additional codebook for collision handling, with T5~\cite{raffel2020exploring} serving as the backbone of the sequence modeling module. the embedding size is 16 and the batch size is 1024. The intent modeling employs a cross-attention mechanism with 2 attention heads and dropout tuned from $\{0.1, 0.2, 0.3\}$. The
MLP structure of the final prediction layer is [256, 256, 256]. 
Model parameters are initialized using Xavier initialization~\cite{glorot2010understanding} and optimized with the Adam optimizer~\cite{diederik2014adam}, with the learning rate of 5e-3. More tuning details and the code are available at \href{https://github.com/RUCAIBox/GenCI}{\textcolor{blue}{https://github.com/RUCAIBox/GenCI}}.

\subsection{Overall Performance}

To validate the effectiveness of GenCI, we compare its performance with several baseline models across three datasets. As summarized in Table~\ref{table:performance}, we make the following observations:

For feature interaction models, traditional methods such as LR and FwFM, which are constrained to modeling low-order feature interactions, generally underperform compared to deep learning-based models such as NFM, PNN and FiGNN that capture higher-order relationships. Ensemble-based approaches like DeepFM, DCNv2, AutoInt+, and xDeepFM further incorporate MLPs to model implicit interactions while also capturing explicit feature interactions. Adaptive models like AFN+ and RFM, which learn arbitrary-order interactions, achieve further gains. Recently, SFG introduces a generative paradigm to refine feature embeddings.
Despite their effectiveness, these models largely follow an embedding-then-feature-interaction paradigm, flattening historical behaviors and failing to capture dynamic user interests.

User behavior modeling approaches mark a paradigm shift by explicitly modeling sequential interactions. DIN leverages target-aware attention to highlight relevant historical behaviors, while DIEN adds an interest evolution layer to capture dynamic interest progression. MIRRN further utilizes multi-granularity queries and Fourier transformers for long-term behavior modeling. By explicitly capturing sequential dynamics, these methods consistently outperform traditional feature interaction baselines.

Ultimately, GenCI achieves state-of-the-art performance across all datasets. By explicitly modeling the NTP task, it captures precise short-term intent that reflects users’ immediate needs. The Hierarchical Candidate-Aware Intent Modeling mechanism further refines this intent into personalized, target-aligned representations.
Finally, a multi-intent fusion module with joint optimization integrates these complementary signals, delivering superior accuracy with inference times comparable to efficient baselines.

\subsection{Experimental Analysis}

\subsubsection{Ablation Study}
To verify the contribution of each key component within our proposed GenCI framework, we conduct a comprehensive ablation study by evaluating several of its variants, as summarized in Table~\ref{table:ablation}.

$\bullet$\textbf{w/o ST-Intent.}
 This variant removes $H_{\text{st}}$ in Eq.~\eqref{equ:pred_v1}, captures refined short-term intent through the \emph{Hierarchical Candidate-Aware Intent Modeling} module. As user behaviors are strongly influenced by immediate interests (discuessed in Section~\ref{sec:TAD interset}), omitting this component results in a significant performance drop, highlighting the importance of modeling dynamic intent at the cohort level.

$\bullet$\textbf{w/o LT-Intent.}
 This variant excludes $H_{\text{lt}}$ in Eq.~\eqref{equ:pred_v1}, which encodes the user’s long-term preference profile captured from historical behavioral sequences. Without this stable global intent representation, the model loses its ability to capture persistent user preferences, leading to inferior overall performance.

$\bullet$\textbf{w/o SR.}
 This variant removes the self-supervised regularization term $\mathcal{L}_{\text{SR}}$ in Eq.~\eqref{equ:joint}, which is designed to stabilize training by providing auxiliary supervision from a baseline CTR model. Without this regularization, the model becomes more sensitive to noisy intent representations generated during the early stages of training, resulting in unstable convergence and degraded accuracy.

 $\bullet$\textbf{w/o NTP.}
 This variant removes the term $\mathcal{L}_{\text{NTP}}$ in Eq.~\eqref{equ:joint}, which is designed to capture sequential behavioral patterns and serve as a recall-oriented objective. Without this supervision, the model fails to align user exploration with final decision-making, directly hindering the achievement of recall-ranking consistency.

Overall, the ablation results clearly demonstrate that each module plays a critical and complementary role in the overall framework. The long-term and short-term intent representations jointly capture the user’s persistent and evolving preferences, while the NTP objective ensures recall-ranking consistency. Furthermore, the self-supervised regularization facilitates stable and reliable training. Together, these components contribute to the superior end-to-end performance of GenCI.

\begin{table}[h]
  % \huge
  \captionsetup{font={small}}
  \caption{Ablation study of the proposed GenCI.}\label{table:ablation}
  \resizebox{.45\textwidth}{!}{
  \begin{tabular}{l|cc|cc|cc}
    \toprule
    \multirow{2}{*}{\textbf{Model}}&
    \multicolumn{2}{c|}{\textbf{MovieLens}}&\multicolumn{2}{c|}{\textbf{Fashion}}&\multicolumn{2}{c}{\textbf{Instruments}}\\
    &AUC&Log Loss& AUC & Log Loss & AUC & Log Loss \\
    \hline\hline
    (0): GenCI & \textbf{0.8964} & \textbf{0.3507} & \textbf{0.7519} & \textbf{0.4959} & \textbf{0.7182} & \textbf{0.3849} \\
    (1): w/o ST-Intent & 0.8935 & 0.3597 & 0.7394 & 0.5211 & 0.7165 & 0.3856 \\
    (2): w/o LT-Intent & 0.8941 & 0.3542 & 0.7414 & 0.6083 & 0.7176 & 0.3858 \\
    (3): w/o SR & 0.8923 & 0.3539 & 0.7410 & 0.5517 & 0.7160 & 0.3856 \\
    (4): w/o NTP & 0.8920 & 0.3550 & 0.7400 & 0.5192 & 0.7146 & 0.3860 \\ 
  \bottomrule
\end{tabular}}
\end{table}

\subsubsection{Performance Comparison \wrt Different Ensemble Approaches}
\begin{figure}[!h]
  \centering
  \captionsetup{font={small}}
  \includegraphics[width=.95\linewidth]{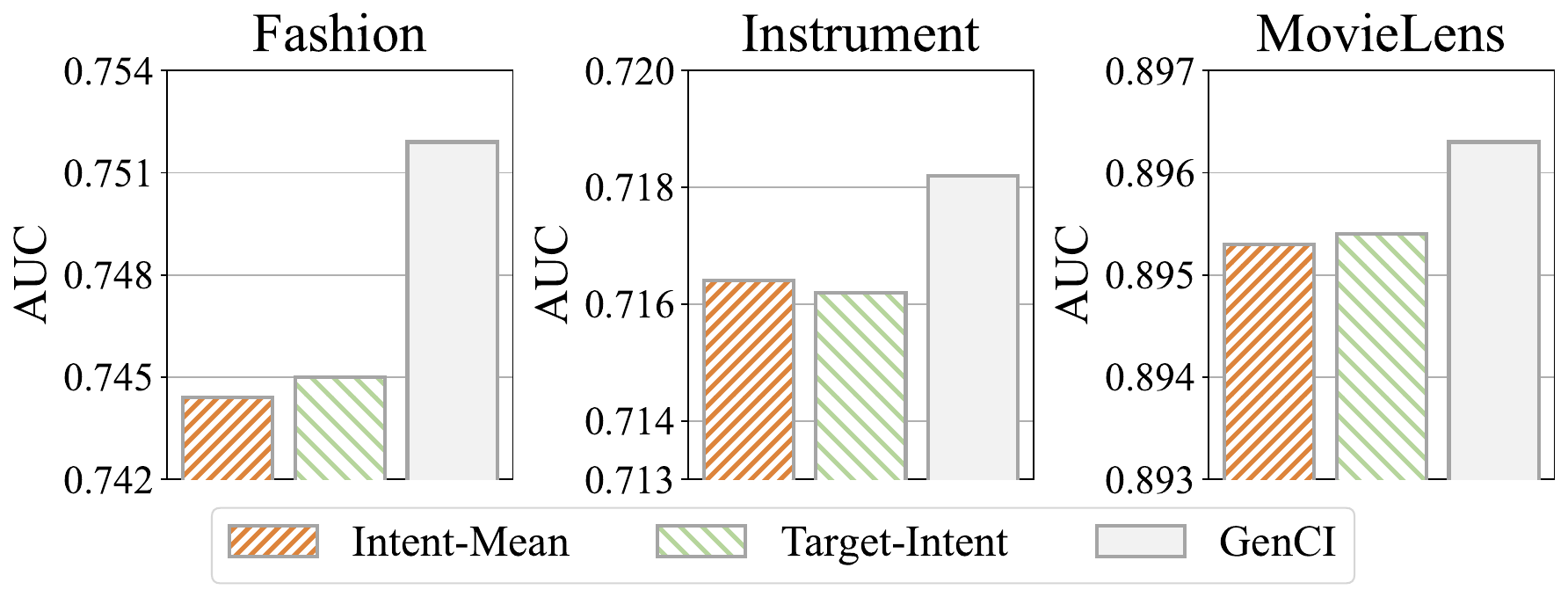}
  % \captionsetup{font={small}}
  \caption{Performance comparison \wrt different ensemble approaches applied in the HCAIM module.}
   \label{fig:albation_ensemle}
\end{figure}

We conducted a set of experiments to evaluate the effectiveness of our HCAIM module under different short-term intent extraction strategies, as shown in Figure~\ref{fig:albation_ensemle}.
\textbf{Intent-Mean} applies mean pooling over cohorts, which treats all cohorts identically and thus neglects the hierarchical item representations learned by RQ-VAE, thereby undermining the preservation of fine-grained intent semantics.
\textbf{Target-Intent} employs target-aware attention to capture target-related representations but overlooks intra-cohort refinement, leaving behaviorally irrelevant items insufficiently filtered.
In contrast, our hierarchical attention mechanism refines cohort representations in a coarse-to-fine manner, enhancing behaviorally relevant and target-aligned intent modeling.

\subsubsection{Visualization of Cohort Interest Distribution}
\begin{figure}[!h]
  \centering
  \captionsetup{font={small}}\includegraphics[width=.85\linewidth]{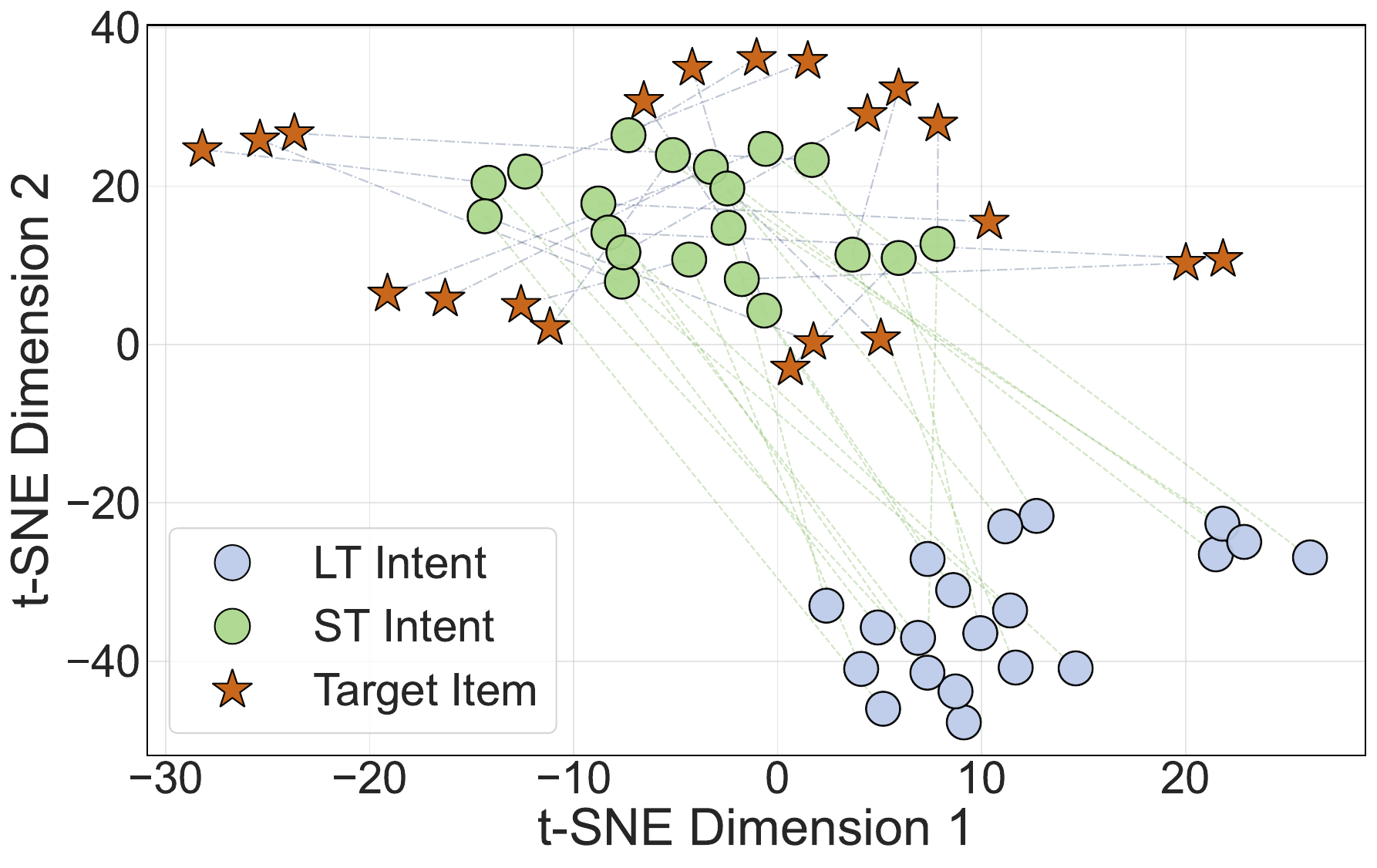}
  % \captionsetup{font={small}}
  \caption{Visualization of different interest distribution. Short-term intents cluster tightly around target items, capturing immediate context, while long-term intents are broadly distributed, representing generalized preferences.}
   \label{fig:vislu}
\end{figure}

To validate the effectiveness of the learned user intents, we employ t-SNE~\cite{maaten2008visualizing} to project the embeddings of long-term and short-term intents, together with target item representations, into a two-dimensional space, as illustrated in Figure~\ref{fig:vislu}.
The short-term intents exhibit compact cohorts centered around their corresponding target items, whereas the long-term intents are more broadly distributed across the latent space.
This clear separation indicates that our model successfully disentangles the two facets of user preference, where the short-term intent captures immediate contextual signals, and the long-term intent encodes stable, generalized user tendencies.

%% file: sec-related.tex
 \section{Related Work}

%  \subsection{Click-Through Rate Models}
The field of CTR prediction has seen significant evolution, with research primarily focused on two areas: feature interaction modeling~\cite{wang2021dcn,guo2017deepfm} and user behavior modeling~\cite{xiao2020deep,zhou2018deep}.

% CTR prediction has evolved along two main directions: feature interaction modeling and user behavior modeling.

\subsection{Feature Interaction Modeling.} Early CTR models such as LR~\cite{richardson2007predicting} are efficient but fail to capture cross-feature signals.
Factorization-machine-based methods~\cite{sun2021fm2,pan2018field,xiao2017attentional,juan2016field} address this by factorizing feature spaces to model second-order interactions under sparse data. The advent of deep learning led to a new paradigm of models capable of learning high-order interactions.
Some models, such as DCNv2~\cite{wang2021dcn} and xDeepFM~\cite{lian2018xdeepfm}, predefine a maximum order to enumerate all feature combinations. To alleviate the limitations of fixed-order designs, other studies~\cite{cheng2020adaptive,tian2024rotative,tian2023eulernet} propose learning the interaction orders adaptively from data, automatically infer interaction orders through transformations in complex vector spaces, achieving greater flexibility.
% Despite their effectiveness, most of these methods treat features as flat inputs and ignore the sequential nature of user behaviors.
While powerful, these models~\cite{li2024dcnv3,li2024tf4ctr,wang2023towards} typically operate on a flattened set of features, often overlooking the inherent sequential structure in user behavior data.

\subsection{User Behavior Modeling.} To overcome the limitations of static feature interaction, subsequent works~\cite{xu2025multi, huang2024recall} model users’ historical behaviors to derive dynamic interest representations.
DIN~\cite{zhou2018deep} introduces an attention mechanism to weight historical behaviors relative to the target item, shifting from static to target-aware profiles.
DIEN\cite{zhou2019deep} integrates GRUs to capture temporal evolution of user interests.
DSIN~\cite{feng2019deep} segments user behaviors into homogeneous sessions and employs LSTMs for finer-grained interest modeling, while BST~\cite{chen2019behavior} leverages the Transformer architecture to capture complex item dependencies. 
Recognizing the importance of long sequences~\cite{si2024twin,chen2021end,cao2022sampling,qin2020user}, models like SIM~\cite{pi2020search} and TWIN~\cite{chang2023twin} adopt two-stage frameworks to effectively process extended user histories. 
However, these methods primarily learn a generalized representation of a user's long-term interests. They are less effective at capturing the immediate, evolving nature of a user's short-term intent, which is critical for predicting their next action. Our work specifically addresses this gap by modeling users’ immediate interests.

%% file: sec-appendix.tex
\clearpage
\newpage

\appendix

% \section{Appendix}
% \input{2_related_work}

\section{Additional Experimental Results}

\subsection{Performance Analysis on Interest Volatility}

To explicitly verify whether GenCI effectively captures dynamic interest shifts, we conducted a simulation experiment by slicing the dataset into static and evolving interest groups. We partitioned users into fast-changing and slow-changing groups based on interest volatility, calculated via the average Jaccard distance of consecutive item genres.

The comparison results in Table~\ref{table:performance_comparison_volalti} show that all models perform worse on the Fast-changing group than on the Slow-changing group, highlighting the challenge of modeling rapid interest shifts. However, GenCI consistently outperforms all baselines, with a more significant improvement in the Fast-changing group. This suggests that while baseline models struggle with quick preference changes, GenCI effectively captures evolving interests using semantic interest cohorts, demonstrating superior robustness in dynamic environments.

\begin{table}[!h]
  % \huge
  \captionsetup{font={small}}
  \caption{Performance comparison on user groups with different interest volatilities.}\label{table:performance_comparison_volalti}
    \resizebox{0.85\columnwidth}{!}{
  \begin{tabular}{l|c c c c}
    \toprule
    \textbf{Model} &
    \textbf{Fast-AUC} & \textbf{Fast-LL} & \textbf{Slow-AUC} & \textbf{Slow-LL} \\
    \hline \hline
    DeepFM & 0.8981  & 0.3653 & 0.9051 & 0.3491 \\
    DCNV2 & 0.8991 & 0.3662 & 0.9052 & 0.3470 \\
    GE4REC & 0.8996 & 0.3725 & 0.9062 & 0.3497 \\
    DIN & 0.8984  & 0.3764 & 0.9056  & 0.3474 \\
    DIEN & 0.8987  & 0.3659 & 0.9086 & 0.3495 \\
    MIRRN & 0.8998  & 0.3691 & 0.9079  & 0.3432 \\
\hline\hline
    GENCI & 0.9049 & 0.3617 & 0.9116 & 0.3431 \\
    \bottomrule
  \end{tabular}}
\end{table}

\subsection{Hyperparameter Analysis}
\begin{figure}[!h]
  \centering
\captionsetup{font={small}}  \includegraphics[width=.9\linewidth]{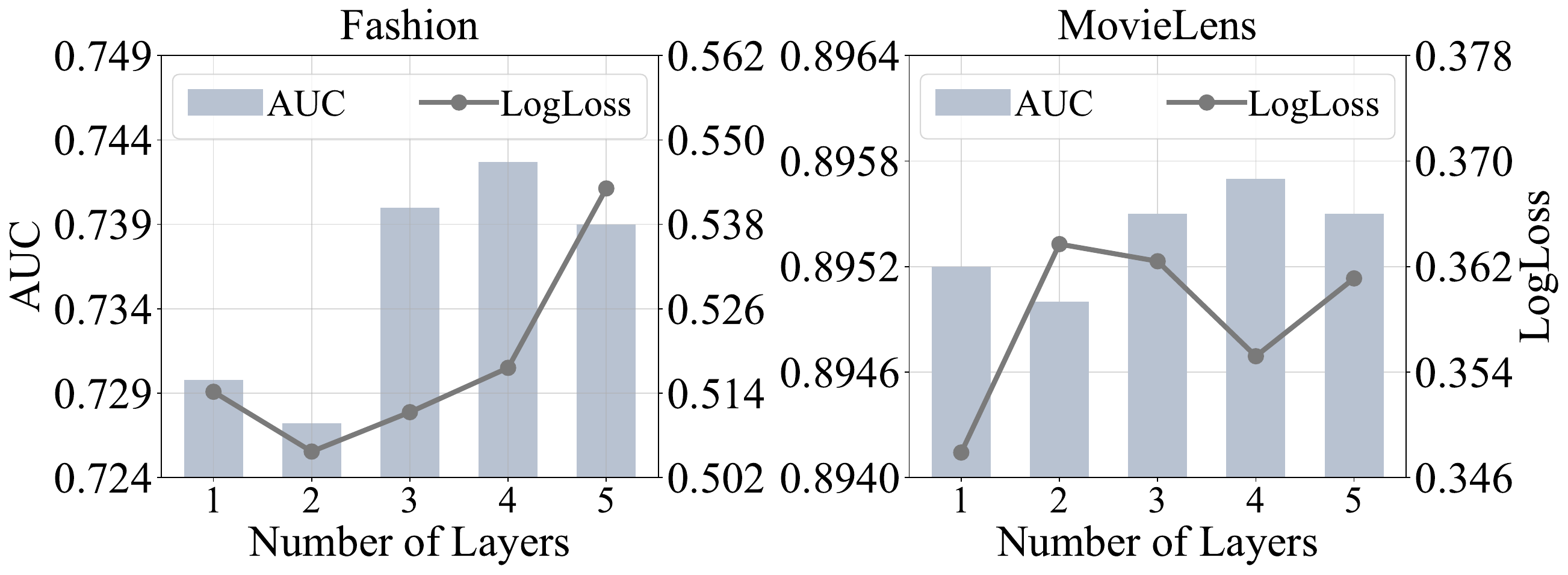}
  % \captionsetup{font={small}}
  \caption{Performance of GenCI \wrt number of Layers}
   \label{fig:hyper_layer}
\end{figure}

We study how the hyperparameters
impact the performance of GenCI, including:

$\bullet$ \emph{Impact of the number of layers}.
GenCI employs $N$ encoder–decoder layers to model the NTP task. To investigate the effect of 
$N$, we vary it from 1 to 5 and report the results in Figure~\ref{fig:hyper_layer}. We observe a clear trend of performance improvement as the number of layers increases, which eventually plateaus beyond a certain depth. This indicates that GenCI can effectively leverage increased model capacity to capture progressively more complex behavioral patterns.

\begin{figure}[!h]
  \centering
\captionsetup{font={small}}  \includegraphics[width=.9\linewidth]{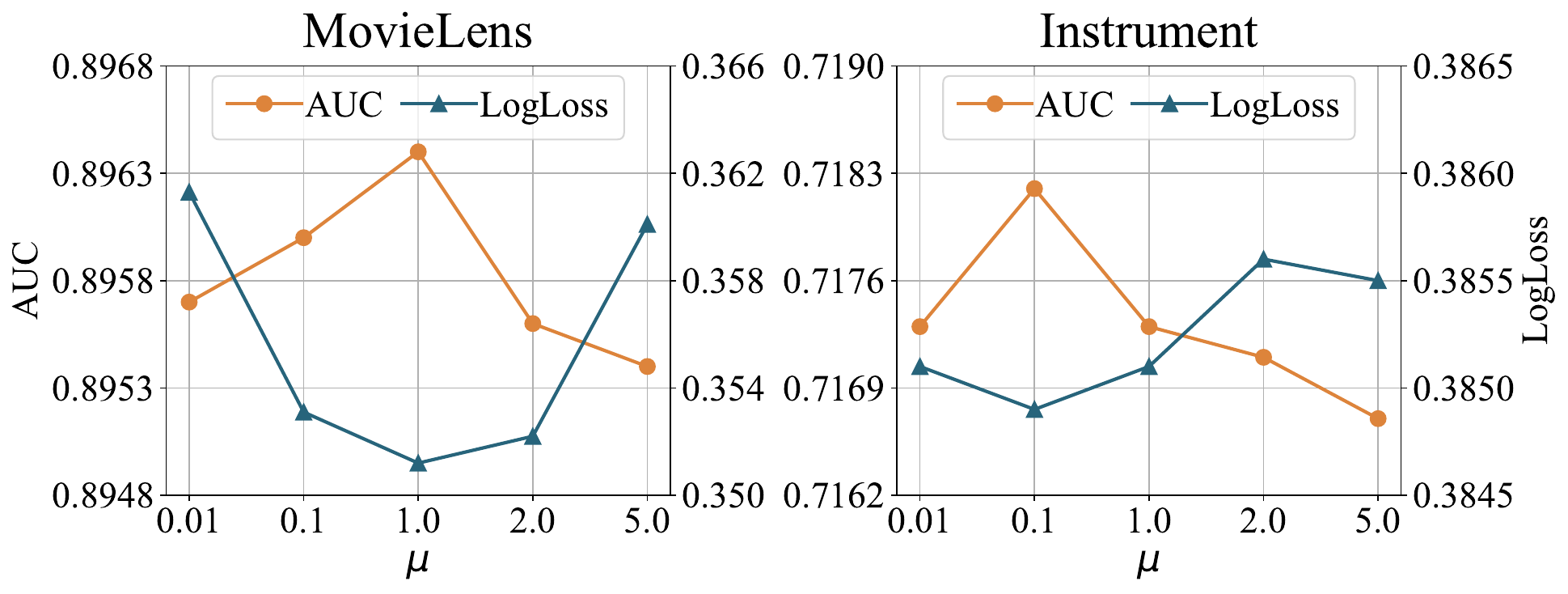}
  % \captionsetup{font={small}}
  \caption{Impact of the coefficients $\mu$ in Loss.}
   \label{fig:mu loss}
\end{figure}

$\bullet$ \emph{Impact of Coefficient $\mu$.} The coefficient $\mu$ in Eq.~\eqref{equ:joint} balances the multi-task objective between the primary CTR task and the auxiliary sequence generation task. We evaluate it over the set $\{0.01, 0.1, 1.0, 2.0, 5.0\}$. Choosing an appropriate value is critical: a small $\mu$ weakens the generative supervision, limiting the model's ability to learn meaningful sequence representations for short-term intents, whereas an excessively large $\mu$ may shift optimization away from the primary CTR objective. As shown in Figure~\ref{fig:mu loss}, datasets with longer average sequence lengths (\eg MovieLens) tend to require larger optimal $\mu$ values to capture more complex temporal dependencies, compared to datasets with shorter sequences (\eg Instruments).

\begin{figure}[!h]
  \centering
\captionsetup{font={small}}  \includegraphics[width=.9\linewidth]{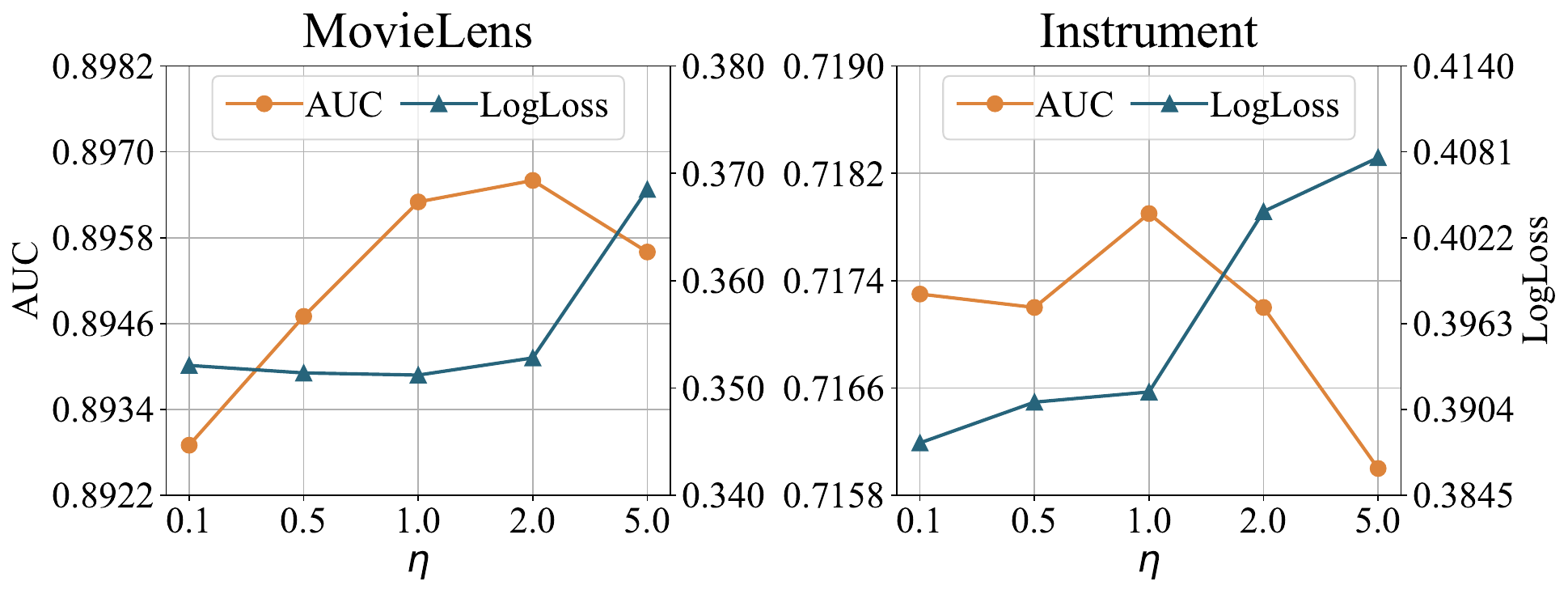}
  % \captionsetup{font={small}}
  \caption{Impact of the coefficients $\eta$ in Loss.}
   \label{fig:eta loss}
\end{figure}

$\bullet$ \emph{Impact of Coefficient $\eta$.} The coefficient $\eta$ in Eq.~\eqref{equ:joint} controls the weight of the self-supervised regularization term. We investigate its sensitivity by varying $\eta$ over the set $\{0.1, 0.5, 1.0, 2.0, 5.0\}$. As shown in Figure~\ref{fig:eta loss}, the performance exhibits a distinct rise-then-fall pattern across both datasets. Specifically, the performance peaks at $\eta=2.0$ for MovieLens and $\eta=1.0$ for Amazon. This suggests that while appropriate regularization facilitates stability, an excessively large $\eta$ may overshadow the primary CTR objective. Furthermore, the higher optimal $\eta$ for MovieLens suggests that feature-rich datasets benefit more from stronger regularization constraints to prevent overfitting.
